\documentclass[12pt]{article}%
\usepackage{amsmath}
\usepackage{amsfonts}
\usepackage{amssymb}
\usepackage{graphicx}%
\providecommand{\U}[1]{\protect\rule{.1in}{.1in}}
\makeatletter
\AtBeginDocument{\@ifpackageloaded{natbib}{\ifNAT@numbers\if@filesw\immediate\write\@auxout{\string\global\string\NAT@numberstrue}\fi\fi}{}}
\makeatother
\begin{document}
\begin{titlepage}
%
\ \\
\begin{center}
\LARGE
{\bf
Black Hole Firewalls Require\\
Huge Energy of Measurement
}
\end{center}
\ \\
\begin{center}
\large{
Masahiro Hotta and Jiro Matsumoto
}\\
{\it
Department of Physics, Faculty of Science, Tohoku University,\\
Sendai 980-8578, Japan
}\\
\ \\
\large{ Ken Funo}\\
{\it
Department of Physics, University of Tokyo,
7-3-1 Hongo, Bunkyo-ku, Tokyo 113-0033, Japan
}
\end{center}
\begin{abstract}
The unitary moving mirror model is one of the best quantum systems
for checking the reasoning of the original firewall paradox of AMPS in quantum black holes.
Though the late-time part of radiations emitted from the mirror is fully entangled with the early-part,
no firewall exists with a deadly, huge average energy flux in this model.
This is because high-energy entanglement structure of the discretized systems in almost maximally entangled states
is modified so as to yield the correct description of low-energy effective field theory. Furthermore, the strong subadditivity paradox of firewalls is resolved
using non-locality of general one-particle states and zero-point fluctuation entanglement.
Due to the Reeh-Schlieder theorem in quantum field theory, another firewall paradox is inevitably raised with quantum remote measurements in the model.
We resolve this paradox from the viewpoint of the energy cost of measurements.
No firewall appears,
as long as the energy for the measurement is much smaller than the ultraviolet cutoff scale.
\end{abstract}
\end{titlepage}

\section{Introduction}

~

The firewall paradox \cite{firewall} of quantum black holes poses a profound
question about the relation between quantum information and quantum gravity.
With the advent of AdS/CFT \cite{AdSCFT}, black hole evaporation processes are
now widely believed to be unitary \cite{hadscft}. It seems likely that all
information stored in the interior of a black hole may be imprinted into
outside radiation and eventually released to the spatial infinity.
Superficially, this process may be accompanied by a quantum-cloning-like
mechanism to generate the informative radiation. However, invoking the concept
of black hole complementarity \cite{bh c} saves the no-cloning theorem
\cite{no c}. Simultaneously, the complementarity maintains the semiclassical
picture in which a free-falling observer experiences nothing out of the
ordinary when crossing the horizon. However, Almheiri et al. (AMPS) recently
argued \cite{firewall} that the absence of drama for the infalling observer
contradicts the following natural postulates as follows:

\bigskip

Postulate 1: The process of formation and evaporation of a black hole, as
viewed by a distant observer, can be described entirely within the context of
standard quantum theory. In particular, there exists a unitary $S$ matrix that
describes the evolution from infalling matter to outgoing Hawking-like radiation.

\bigskip

Postulate 2: Outside the stretched horizon of a massive black hole, physics
can be described to a good approximation by a set of semiclassical field equations.

\bigskip

Assuming Postulates 1 and 2, AMPS analyze the entanglement ability of gaining
information about late-time radiation for a sufficiently old black hole via
measurements of early-time radiation. They then conclude that infalling
observers will encounter high-average-energy modes of the late-time radiation
and burn up at the horizon. Because of Postulate 1, the final state
$|\Psi\rangle$ of black hole evaporation is assumed to be pure. They divide
the system into an early radiation part $E$ and a late radiation part $L$. It
is assumed that there is a natural UV cutoff in the black hole physics and
that the Hilbert spaces of $E$ and $L$ have finite discrete dimensions
$N_{E}\,$and$\ N_{L}$. Using an arbitrary complete basis $\left\{
|i\rangle_{L}\right\}  $ for $L~$of an old black hole with $N_{E}>N_{L}$, the
final state $|\Psi\rangle$ can be expanded as:
\begin{equation}
|\Psi\rangle=\sum_{i=1}^{N_{L}}|\psi_{i}\rangle_{E}|i\rangle_{L}, \label{01}%
\end{equation}
where $|\psi_{i}\rangle_{E}$ is an unnormalized state of $E$. Using the same
philosophy of typical-state entanglement \cite{l,LP,page}, one can argue that
$L$ is fully entangled with $E$ in the state $|\Psi\rangle$. Thus, AMPS assume
that a measurement of $E$ outputting a result $i$ exists such that the
post-measurement state of $L$ is nearly equal to an arbitrarily fixed
$|i\rangle_{L}$. In particular, a measurement of the number of outgoing $L$
particles may be expected, in which$\ |i\rangle_{L}$ becomes a number
eigenstate of the particles. Going back in time, the particles in state
$|i\rangle_{L}$ are present near the horizon and severely blueshifted by the
strong gravitational force. Each particle carries quite a huge amount of
energy---much larger than the Hawking temperature of the black hole. Thus AMPS
expects that the energy expectation value of the particles also becomes
divergent. This high-average-energy flux of the mode near the horizon is
referred to as a \textit{firewall}. It is also stressed that the real
execution of the measurement is not needed for firewalls \cite{MP}\cite{AS2}.
In the argument, the existence of almost maximal entanglement between $E$ and
$L$ is essential. Because the reduced density operator of $L$ becomes an
almost maximally entropic state,%
\[
\operatorname*{Tr}_{E}\left[  |\Psi\rangle\langle\Psi|\right]  \sim\frac
{1}{N_{L}}\sum_{i}|i\rangle_{L}\langle i|_{L}=\frac{1}{N_{L}}\mathbf{I}_{L}%
\]
it is concluded that typical states with firewall singularity are dominant in
the calculation of expectation values of the particle number and the energy
momentum tensor. Let us imagine Alice in a free-fall motion tries to get
across the horizon. Then, irrespective of whether the firewall observation
requires a measurement of $E$, she will encounter the firewall with near
certainty before she passes over the horizon and burns out, as long as AMPS's
argument is correct. This firewall scenario is completely different from the
complementarity scenario \cite{bh c}, in which Alice is able to safely
traverse the horizon.

Exactly as in Ref. \cite{mathur}, AMPS also restate this firewall conjecture
from an information theoretical point of view. Let $A$ be the early radiation
modes; $B$, the late radiation modes; and $C$, the modes inside the horizon.
AMPS argue that the strong subadditivity relation of the entropy \cite{nc},%
\begin{equation}
S_{AB}+S_{BC}\geq S_{B}+S_{ABC}, \label{1}%
\end{equation}
is violated unless a free-falling observer burns out by the firewall. They say
that the absence of tragic drama implies $S_{BC}=0$ and so $S_{ABC}=S_{A}$.
Because $S_{B}>0$ is guaranteed by the thermality of $B$, $S_{AB}>S_{A}$ is
derived from Eq. (\ref{1}). In fact, the black hole loses its entropy and
$S_{AB}<S_{A}$ holds. Therefore, they conclude that the no-firewall assumption
is wrong. Besides, since $A$ and $B$ are almost maximally entangled with each
other in a typical state by the same philosophy of Page curve \cite{pagetime},
$B$ is purified by a subsystem of $A$, which is denoted by $\bar{B}_{A}$
($\subset A$) \cite{HH}. Thus it seems that $B$ and $\bar{B}_{A}$ are in a
pure entangled state. This implies $S_{\bar{B}_{A}B}=0$ and $S_{\bar{B}_{A}%
BC}=S_{C}$. Therefore the typical-state argument and strong additivity among
$\bar{B}_{A},B$ and $C$ derive $I_{BC}=S_{B}+S_{C}-S_{BC}\leq0$. Because
$I_{BC}$ is mutual information of $B$ and $C$, $I_{BC}$ is non-negative. Thus
we obtain $I_{BC}=0$. However, the no-drama condition implies $I_{BC}\gg1$ and
seems to contradict the typical-state result.

Only a very short while after this paradox was raised last year, there have
already been numerous efforts \cite{reffw} to resolve it by direct attacks to
the gravitational system. However, the answer remains elusive because we are
yet to arrive at a quantum gravity theory. In this paper, a different strategy
is taken against the paradox. We analyze not the gravitational system but
instead a moving mirror model of a free massless scalar field as one of the
best quantum systems to test the validity of AMPS's reasoning about the
firewall It is well known that moving mirrors reproduce emission of Hawking
radiation \cite{fd,BD}. The dynamics of the system is very simple and
completely unitary. In accordance with the Reeh--Schlieder theorem
\cite{rs,PT}, $L$ is fully entangled with $E$. Based on the theorem, it can be
argued that any state of $L$ can be arbitrarily reproduced closely by
operating a polynomial of local $E$ operators on the final state of scattering
with the mirror. Owing to the finiteness of the Hilbert space dimensions, this
means that we can construct a local measurement operator of $E$ \cite{nc} that
yields an arbitrary post-measurement state of $L$. \ 

Though the AMPS assumptions seem satisfied on first glance, it is noticed that
no firewall appears with deadly huge average energy flux in the moving mirror
model, because the average values of energy-momentum tensor are finite
everywhere. As argued in Section 3, the reason of no firewalls is that the
continuum limit using regularization with Poincar\'{e} and conformal
invariances modifies the UV entanglement structure of typical states to
generate a low-energy effective field theory. Therefore, the maximal
entanglement condition assumed by AMPS is not sustained in this limit. It
seems very plausible that this modification of the entanglement structure in
the continuum limit may also occur for the entanglement of $E$ and $L$ of the
gravitational system to avoid the original AMPS paradox. In this paper, it is
also argued that the strong subadditivity argument of AMPS has a serious flaw
from a viewpoint of locality.

Our results suggest that Page curve argument is not able to be applied
precisely to black hole evaporation. Originally, the argument is based on
appearance of almost maximum entanglement in typical-state models with huge
degeneracy and small interaction \cite{l,LP,page}. The models are proposed
just for exploring foundation of statistical mechanics in macroscopic systems
which consist of a huge number of the same-energy components interacting with
each other via very small coupling constants. However, many interacting
systems like ordinary (uniformly distributed) spin networks with finite
dimensions do not satisfy this condition. The number density of states usually
increases very fast as energy increases. Thus the energy of "typical states"
in the Hilbert space is of almost the same order of the maximum energy of the
system. If we ignore the low-energy-state contribution, the standard analyses
\cite{l,LP,page} mean that two large complementary subsystems have almost
maximum entanglement in a typical state with typical energy, though the energy
is very high. Therefore, entanglement entropy between the subsystems is
proportional to volume of the smaller subsystem if the subsystems are
uniformly distributed. This is actually a volume law of entanglement entropy,
not an area law. As opposed to the typical states with typical high energy, it
is well known that low-energy states near the ground state obey area laws of
entanglement entropy \cite{S}\cite{cirac}. Note that, even in quantum field
theory, entanglement entropy of the vacuum state and low-energy states also
obeys the (approximated) area law, not the volume law \cite{S}. This may imply
that the almost maximum entanglement cannot be attained in ordinary low-energy
states of quantum fields. Hence, in the context of firewall arguements, it is
naturally expected that the description of low-energy field theory does not
allow the almost maximum entanglement between two complemetary subsystems.
Thus the Page curve picture may be inappropriate in cases with initial
low-energy states, which are prepared by some physical selection rule like
\cite{ecc}.

\bigskip

Although we do not have any firewall with nonzero expectation values of
firewall particle number, an extended firewall paradox inevitably arises in
this model when a general measurement of $E$ is performed, as shown in Section
4. (The similar paradox appears in a more simple case of Rindler horizon.)
However, we prove that the measurement does not make the firewall emerge,
provided the energy cost of the $E$ measurement, which outputs information
about $L$, is much smaller than the ultraviolet cutoff scale. In the black
hole system, a similar paradox may arise, but too much measurement energy
induces a large back reaction to spacetime and may cause formation of a new
black hole in the measurement region and enclose the measurement device within
the event horizon before it outputs results. Thus a conjecture, firewall
information censorship (FIC), can be proposed whereby information about
encounters with firewalls is never exposed to our low-energy world. This is a
conjecture similar to that of extreme cosmic censorship (ECC) \cite{ecc},
which selects out regular initial states to avoid firewalls. However, it
should be emphasized that ECC does not prohibit the firewall emergence when we
perform general $E$ measurements. In this paper, it is also argued that vacuum
fluctuation is fully entangled with the late radiation modes $B$ and the
interior modes $C$. This implies that $S_{BC}>0$ without generating the
firewall, and it avoids the violation of strong subadditivity in Eq. (\ref{1}).

The paper is organized as follows. In Section 2, the firewall paradox is posed
along the line of reasoning of \cite{firewall} in a moving mirror model. The
resolution of the paradox is described in Section 3. The entropic firewall
paradox regarding strong subadditivity is resolved from the viewpoint of
strictly localized states in quantum field theory. In Section 4, we pose
another firewall paradox based on quantum measurement. The paradox is resolved
from a viewpoint of energy cost of measurements. This results suggests a
cosmic censorship conjecture in quantum information theory. In Section 5, we
summarize our results. We adopt the natural units $c=\hbar=1$ in this paper.

\bigskip

\bigskip

\section{AMPS Paradox for a Moving Mirror}

\bigskip

~

\begin{figure}[t]
\begin{center}
\includegraphics[height=.5\textheight]{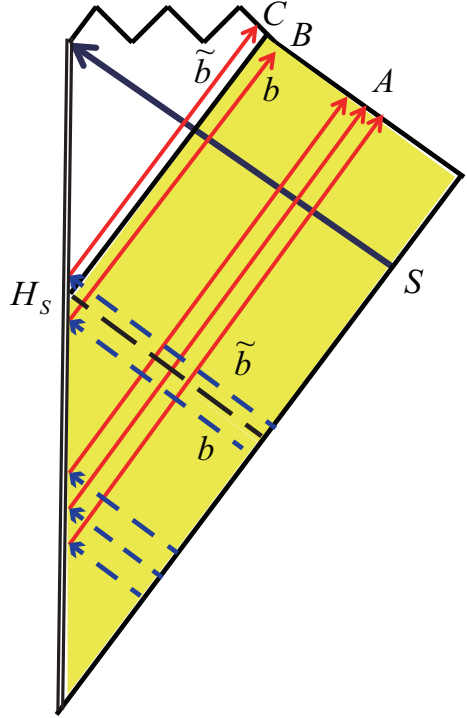}
\end{center}
\caption{(color online). Penrose diagram of the gravitational collapse of a
massless spherical shell.}%
\label{fig1}%
\end{figure}

\begin{figure}[t]
\begin{center}
\includegraphics[height=.5\textheight]{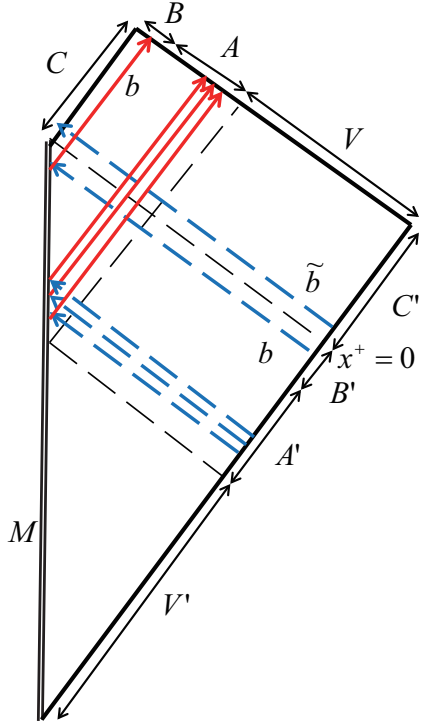}
\end{center}
\caption{(color online). Schematic diagram of a moving mirror with the
trajectory in Eq. (\ref{3}).}%
\label{fig2}%
\end{figure}

It is known that by omitting the curvature outside event horizons, moving
mirror models can mimic the gravitational collapse of a spherical massless
shell in Einstein gravity \cite{w,hlw}. The mirror arises at the origin of the
spherical coordinates ($r=0$). The mirror motion is caused by the time
evolution of the global spacetime structure. Fig.~\ref{fig1} depicts a
time-radius Penrose diagram of the collapse. The arrow denoted by $S$
indicates the massless shell motion, which forms an event horizon $H_{S}$.
Entangled Rindler modes at both sides of $H_{S}$ are denoted by $b$ and
$\tilde{b}$ in Fig.~\ref{fig1} and are fully entangled. The shaded region in
Fig.~\ref{fig1} can be mapped into the moving-mirror flat spacetime, which is
depicted as a truncated Penrose diagram in Fig.~\ref{fig2}. $M$ in
Fig.~\ref{fig2} denotes the mirror trajectory, corresponding to $r=0$ of the
gravitational collapse. The horizon $H_{S}$ is mapped into $x^{-}=\infty$ in
this diagram. Clearly, in the moving mirror model, $\tilde{b}$ of $C$
corresponds to a late-time infalling mode.

Let us consider a (1+1)-dimensional flat spacetime with a moving mirror, the
trajectory of which is given by%
\[
x^{+}=f\left(  x^{-}\right)  ,
\]
where $x^{\pm}$ are light-cone coordinates defined by $x^{\pm}=t\pm x$
\cite{fd,BD}. On the right-hand-side region, a massless scalar quantum field
$\hat{\varphi}$ obeys the equation of motion as $\left(  \partial_{t}%
^{2}-\partial_{x}^{2}\right)  \hat{\varphi}=0$, and it vanishes at the
location of the mirror:
\[
\hat{\varphi}|_{x^{+}=f\left(  x^{-}\right)  }=0.
\]

\begin{figure}[t]
\begin{center}
\includegraphics[height=.4\textheight]{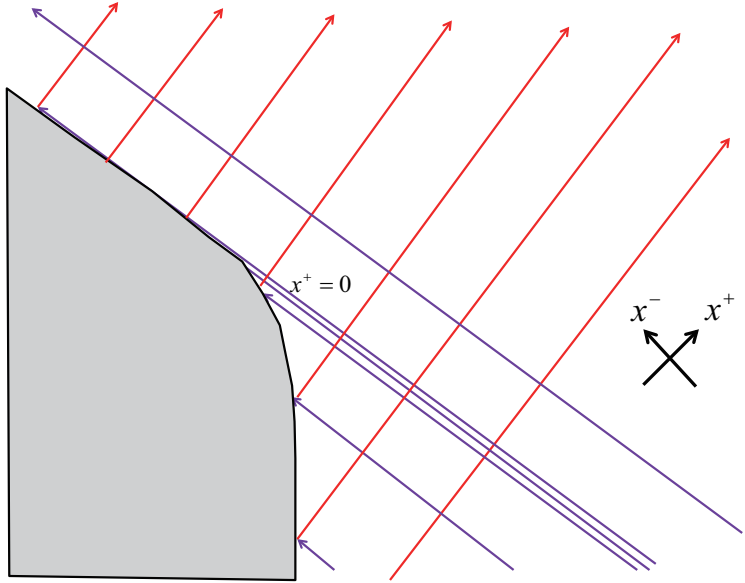}
\end{center}
\caption{(color online). Penrose diagram of a moving mirror with the
trajectory in Eq. (\ref{3}).}%
\label{fig3}%
\end{figure}

The solution is given by%
\begin{equation}
\hat{\varphi}=\hat{\varphi}_{in}\left(  x^{+}\right)  -\hat{\varphi}%
_{in}\left(  f\left(  x^{-}\right)  \right)  , \label{2}%
\end{equation}
where $\hat{\varphi}_{in}(x^{+})$ is the incoming field operator.
$\hat{\varphi}_{in}(x^{+})$ can be expanded as%

\[
\hat{\varphi}_{in}(x^{+})=\int_{0}^{\infty}\left(  \hat{a}_{\omega}e^{-i\omega
x^{+}}+\hat{a}_{\omega}^{\dag}e^{i\omega x^{+}}\right)  \frac{d\omega}%
{\sqrt{4\pi\omega}}%
\]
with creation and annihilation operators $\hat{a}_{\omega}^{\dag}$ and
$\hat{a}_{\omega}$ satisfying $\left[  \hat{a}_{\omega},\hat{a}_{\omega
^{\prime}}^{\dag}\right]  =\delta\left(  \omega-\omega^{\prime}\right)  $. The
in-vacuum state $|0_{in}\rangle$ is defined by $\hat{a}_{\omega}|0_{in}%
\rangle=0$. The outgoing field operator $\hat{\varphi}_{out}\left(
x^{-}\right)  $ is introduced by
\[
\hat{\varphi}_{out}\left(  x^{-}\right)  =\hat{\varphi}_{in}\left(  f\left(
x^{-}\right)  \right)  .
\]
In terms of plane-wave modes, $\hat{\varphi}_{out}\left(  x^{-}\right)  $ can
be expanded as
\[
\hat{\varphi}_{out}(x^{-})=\int_{0}^{\infty}\left(  \hat{b}_{\omega
}e^{-i\omega x^{-}}+\hat{b}_{\omega}^{\dag}e^{i\omega x^{-}}\right)
\frac{d\omega}{\sqrt{4\pi\omega}},
\]
where $\hat{b}_{\omega}^{\dag}$ and $\hat{b}_{\omega}$ are creation and
annihilation operators obeying $\left[  \hat{b}_{\omega},\hat{b}%
_{\omega^{\prime}}^{\dag}\right]  =\delta\left(  \omega-\omega^{\prime
}\right)  $. Unless the mirror is in inertial motion, $\hat{b}_{\omega}~$is
described as a mixture of $\hat{a}_{\omega}$ and $\hat{a}_{\omega}^{\dag}$
such that
\[
\hat{b}_{\omega}=\int_{0}^{\infty}d\omega^{\prime}\left(  B_{\omega}%
(\omega^{\prime})\hat{a}_{\omega^{\prime}}+C_{\omega}(\omega^{\prime})\hat
{a}_{\omega^{\prime}}^{\dag}\right)  .
\]
The appearance of the nonvanishing coefficient $C_{\omega}(\omega^{\prime})$
means particle creation induced by the mirror motion. Let us first consider
the mirror trajectory
\begin{equation}
f(x^{-})=-\frac{1}{\kappa}\ln\left(  1+e^{-\kappa x^{-}}\right)  , \label{3}%
\end{equation}
where $\kappa$ is a positive parameter. The mirror stops in the far past
($x^{-}\sim-\infty$) and approaches a light geodesic $x^{+}=0$ in the future
($x^{-}\sim\infty$). In Fig.~\ref{fig3}, the trajectory is drawn in spacetime.
The left-going lines represent incoming light rays and the right-going lines
represent the reflective light rays. By solving the mirror trajectory as
$x^{-}=g(x^{+})$, the outgoing mode function is described as

\[
v_{\omega}(x^{+})=e^{-i\omega g(x^{+})}=\left(  e^{-\kappa x^{+}}-1\right)
^{i\frac{\omega}{\kappa}},
\]
where $-\infty<x^{+}<0$. In a way similar to that used for computing
$C_{\omega}(\omega^{\prime})$ in the original work by Hawking \cite{h},
approximating $v_{\omega}(x^{+})$ by its dominant contribution $\left(
-\kappa x^{+}\right)  ^{i\frac{\omega}{\kappa}}$ around $x^{+}\sim0$ yields
\[
\langle0_{in}|\hat{b}_{\omega}^{\dag}\hat{b}_{\omega}|0_{in}\rangle=\int
_{0}^{\infty}\left\vert C_{\omega}(\omega^{\prime})\right\vert ^{2}%
d\omega^{\prime}\varpropto\frac{1}{\exp(\frac{2\pi}{\kappa}\omega)-1}.
\]
This implies that the mirror generates thermal radiation with temperature
$T=\kappa/(2\pi)$. This interpretation can be cross-checked by estimating the
energy flux. In general, a mirror with a trajectory $x^{+}=f(x^{-})$ emits
energy flux given by%
\begin{equation}
\langle0_{in}|\hat{T}_{--}(x^{-})|0_{in}\rangle=-\frac{1}{24\pi}\left[
\frac{\partial_{x^{-}}^{3}f(x^{-})}{\partial_{x^{-}}f(x^{-})}-\frac{3}%
{2}\left(  \frac{\partial_{x^{-}}^{2}f(x^{-})}{\partial_{x^{-}}f(x^{-}%
)}\right)  ^{2}\right]  . \label{4}%
\end{equation}
For the trajectory in Eq. (\ref{3}), the energy flux at $x^{-}\gg1/\kappa$
coincides with the thermal flux with temperature $T$:
\[
\langle0_{in}|\hat{T}_{--}(x^{-}\gg1/\kappa)|0_{in}\rangle=\frac{\pi}{12}%
T^{2}.
\]

At the past null infinity of Fig.~\ref{fig2}, the field $\hat{\varphi}$ is in
$|0_{in}\rangle$. The system at the past null infinity is divided into three
subsystems: $V^{\prime}$, $C^{\prime}$, and a composite system $A^{\prime
}B^{\prime}$. The systems $V^{\prime},$ $A^{\prime}$, and $B^{\prime}$ evolve
into $V$, $A$, and $B$ of the future null infinity with $x^{+}\sim\infty$. $A$
corresponds to the early radiation modes and $B$ to the late radiation modes
in the gravitational collapse in Fig.~\ref{fig1}. Subsystem $V$ denotes the
out-vacuum fluctuation with no radiation energy. The composite system $AB$
denotes the radiation emitted from the mirror. The red lines in
Fig.~\ref{fig2} represent the rays of radiation. The broken lines represent
the infalling modes corresponding to the rays. Subsystem $C^{\prime}$ evolves
into $C$ of a separate future null infinity with $x^{-}\sim\infty$. Subsystem
$C$ corresponds to the mode absorbed by a black hole, that is, the interior
mode, as mentioned above.

\begin{figure}[t]
\begin{center}
\includegraphics[height=.4\textheight]{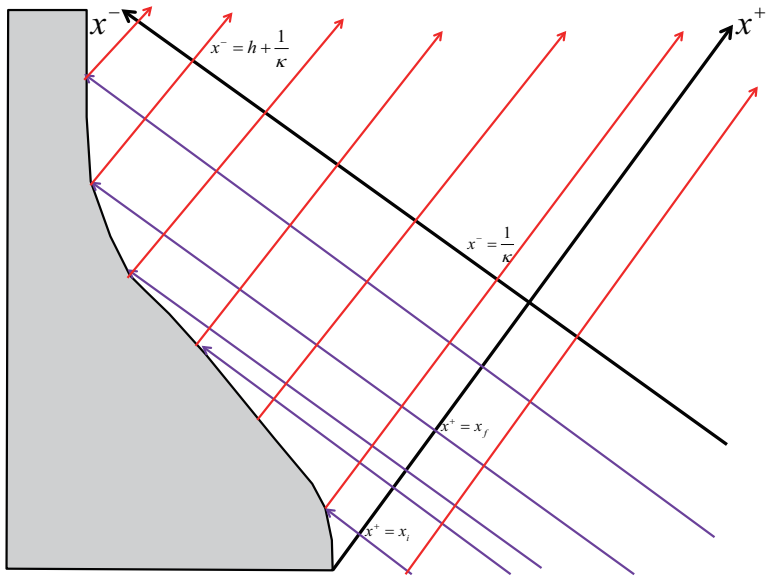}
\end{center}
\caption{(color online). Schematic diagram of a moving mirror with the
trajectory in Eq. (\ref{10}).}%
\label{fig4}%
\end{figure}

\begin{figure}[t]
\begin{center}
\includegraphics[height=.5\textheight]{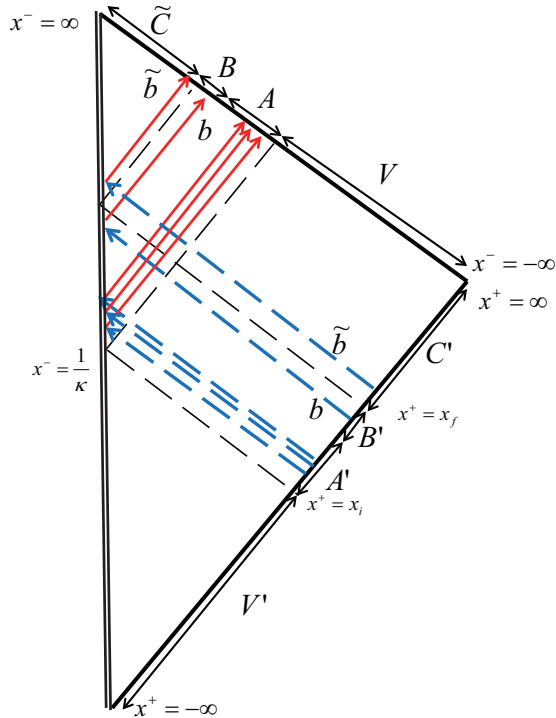}
\end{center}
\caption{(color online). Penrose diagram of a moving mirror with the
trajectory in Eq. (\ref{10}).}%
\label{fig5}%
\end{figure}

Now let us formulate a firewall paradox in this model. Consider a mirror
trajectory given by
\begin{equation}
f_{h}(x^{-})=-\frac{1}{\kappa}\ln\left(  \frac{1+e^{-\kappa x^{-}}%
}{1+e^{\kappa\left(  x^{-}-h\right)  }}\right)  , \label{10}%
\end{equation}
where $h$ is a real parameter satisfying $h\gg1/\kappa$. When $h\rightarrow
\infty$, the mirror trajectory approaches the trajectory in Eq. (\ref{3}).
However, the future structure of spacetime is different. The trajectory is
depicted in Fig.~\ref{fig4}. At $x^{-}\gg h+1/\kappa$, the mirror comes to
rest and eventually stops emitting radiation. The Penrose diagram is shown in
Fig.~\ref{fig5}. As opposed to the situation in Fig.~\ref{fig2}, we have a
single future null infinity. The region of the almost thermal radiation
subsystem $AB$ is now truncated to $\left(  1/\kappa,h+1/\kappa\right)  $. As
shown in Fig.~\ref{fig5}, subsystem $\tilde{C}$ mimics the final informative
radiation in complete evaporation of a black hole. $\tilde{C}$ is the
entangled partner of $ABV$ and purifies the total system in this model.
Because it is still unclear whether the moving mirror model has a holographic
description like AdS, we do not know whether $\tilde{C}$ really corresponds to
a subsystem of a boundary CFT. However, even if this is the case, it should be
stressed that $\tilde{C}$ in our model does not correspond to all the interior
modes inside the black hole including the initial collapsing matter.
$\tilde{C}$ merely carries information of the free field interior modes
$\hat{\varphi}$ to guarantee that the final state of the $\hat{\varphi}$ field
is exactly pure. Hence, our argument does not conclude that $A=R_{B}$ or
$\tilde{B}$ $\subset E$, as discussed in literature \cite{A=E}.

\begin{figure}[t]
\begin{center}
\includegraphics[height=.4\textheight]{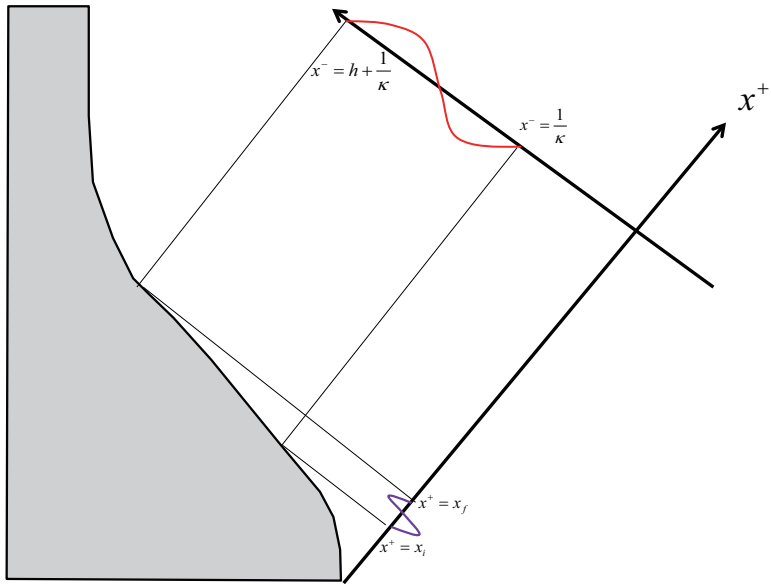}
\end{center}
\caption{(color online). Schematic diagram representing the blueshift
(redshift) by the moving mirror.}%
\label{fig6}%
\end{figure}

If we have a mode excitation of $AB$ at the future null infinity, depicted as
the red curve in Fig.~\ref{fig6}, the mode is strongly blueshifted before
scattering with the mirror and returns in time to a region defined by $\left(
x_{i},x_{f}\right)  $ at the past null infinity, as shown by the
short-interval wave curve in Fig.~\ref{fig6}. The severe blueshift occurs
because the width of the past subsystem $A^{\prime}B^{\prime}$ is much smaller
than that of the future subsystem $AB$:
\[
x_{f}-x_{i}\ll h.
\]

\begin{figure}[t]
\begin{center}
\includegraphics[height=.5\textheight]{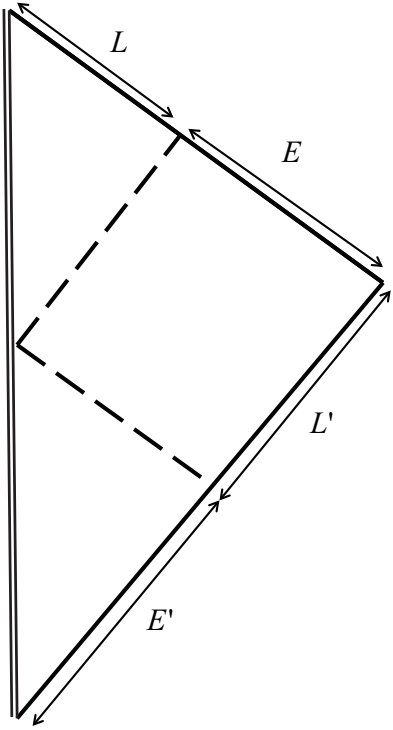}
\end{center}
\caption{Penrose diagram representing the division setup of the system at
asymptotic null infinities. The broken line denotes the boundary of $E$ and
$L$.}%
\label{fig7}%
\end{figure}

This classical blueshift property plays an essential role in the paradox. Note
that any entanglement is conserved in time because the modes are simply
stretched during the reflection by the mirror. Thus, we can always analyze
entanglement at a future time by using the past entanglement. Based on this
fact, it can be argued that a late-mode subsystem $L$, which is a composite
system of $B$ and $\tilde{C}$, is fully entangled with an early-mode subsystem
$E$ composed of $V$ and $A$. As opposed to AMPS, it should be stressed that
$L$ includes the future zero-point fluctuation after the mirror stops and that
$E$ includes the past zero-point fluctuation before the mirror moves. \ In
Fig.~\ref{fig7}, a subsystem\ $L^{\prime}$ ($E^{\prime}$) at the past null
infinity time evolves to $L$ ($E$). Therefore, entanglement between $L$ and
$E$ is equal to that between $L^{\prime}$ and $E^{\prime}$ in $|0_{in}\rangle
$. Using Unruh modes \cite{Unruh}, it turns out that $L^{\prime}$ and
$E^{\prime}$ are fully entangled such that%
\begin{equation}
|0_{in}\rangle=%
{\displaystyle\prod\limits_{\omega}}
\left(  \sum_{n}\frac{e^{-n\frac{\pi\omega}{a}}}{\sqrt{Z(\frac{\omega}{a})}%
}|n(\omega)\rangle_{L^{\prime}}|n(\omega)\rangle_{E^{\prime}}\right)  ,
\label{6}%
\end{equation}
where $a$ is a free positive parameter that indicates the acceleration of the
Unruh modes, $Z(\frac{\omega}{a})=\sum_{n}e^{-n\frac{2\pi\omega}{a}}$, and
$|n(\omega)\rangle_{L^{\prime}}$($|n(\omega)\rangle_{E^{\prime}}$) is the
number eigenstate of $n$ Rindler particles with frequency $\omega$ for
$L^{\prime}$ ($E^{\prime}$). If the modes are regularized in terms of the
Unruh modes with frequency cutoff $K$ and particle number cutoff $N$, it is
directly verified that $|0_{in}\rangle$ is indeed an almost maximally
entangled state between $E^{\prime}$ and $L^{\prime}$ when we consider a large
$a$:
\begin{align}
|0_{in}\rangle &  =\lim_{a\rightarrow\infty}%
{\displaystyle\prod\limits_{k=1}^{K}}
\left(  \sum_{n=0}^{N}\frac{e^{-n\frac{\pi\omega_{k}}{a}}}{\sqrt
{Z(\frac{\omega}{a})}}|n(\omega_{k})\rangle_{L^{\prime}}|n(\omega_{k}%
)\rangle_{E^{\prime}}\right) \nonumber\\
&  =%
{\displaystyle\prod\limits_{k=1}^{K}}
\left(  \frac{1}{\sqrt{N}}\sum_{n=0}^{N}|n(\omega_{k})\rangle_{L^{\prime}%
}|n(\omega_{k})\rangle_{E^{\prime}}\right)  . \label{e3}%
\end{align}
Therefore, in this regularization, $|0_{in}\rangle~$can be regarded as a
typically entangled state in the context of References \cite{l,LP,page}. Of
course, the same is true for $E$ and $L$ at the future null infinity. Thus
this model satisfies the entanglement condition of Reference \cite{firewall}.
However, it should be emphasized here that no firewalls with divergent
expectation values of the energy--momentum tensor appear in the model. In
fact, the expectation value of the energy--momentum tensor is finite
everywhere. Why does the firewall disappear? The reasons are explained in the
next section.

\section{Resolution of AMPS Paradox}

The regularization, which is adopted in the previous section, satisfies
neither special relativistic invariance nor conformal invariance. In
particular, it does not satisfy transitional invariance under $x^{+}%
\rightarrow x^{+}+c$. In fact, the state of the border region between
$E^{\prime}$ and $L^{\prime}$ ($E$ and $L$) becomes singular when we consider
$K\rightarrow\infty$ and $N\rightarrow\infty$. This implies that the
regularization does not allow a continuum limit to a low-energy field theory.
If we adopt any other regularization that maintains the conformal invariance,
the superficial $a$ dependence in Eq. (\ref{6}) is eliminated. More precisely,
the Unruh representation of $|0_{in}\rangle$ is given by:
\[
|0_{in}\rangle\varpropto\exp\left(  \int_{0}^{\infty}e^{-\frac{\pi\omega}{a}%
}b_{\omega}^{\dag}\tilde{b}_{\omega}^{\dag}d\omega\right)  |0_{Rindler}%
\rangle,
\]
where $b_{\omega}^{\dag}$ and $\tilde{b}_{\omega}^{\dag}$ are creation
operators of the Rindler particles satisfying $\left[  b_{\omega_{1}%
},~b_{\omega_{2}}^{\dag}\right]  =\left[  \tilde{b}_{\omega_{1}},~\tilde
{b}_{\omega_{2}}^{\dag}\right]  =\delta\left(  \omega_{1}-\omega_{2}\right)
$, and $|0_{Rindler}\rangle$ is the Rindler vacuum state defined by
$b_{\omega}|0_{Rindler}\rangle=\tilde{b}_{\omega}|0_{Rindler}\rangle=0$. By
changing variables as $\omega^{\prime}=\frac{\omega}{a},~b_{\omega^{\prime}%
}^{\prime}=\sqrt{a}b_{\omega},\text{ and}~\tilde{b}_{\omega^{\prime}}^{\prime
}=\sqrt{a}\tilde{b}_{\omega}$, it is easily checked that the $a$ dependence
vanishes in $|0_{in}\rangle$ as%
\begin{equation}
|0_{in}\rangle\varpropto\exp\left(  \int_{0}^{\infty}e^{-\pi\omega^{\prime}%
}b_{\omega^{\prime}}^{\prime\dag}\tilde{b}_{\omega^{\prime}}^{\prime\dag
}d\omega^{\prime}\right)  |0_{Rindler}\rangle, \label{06}%
\end{equation}
where $\left[  b_{\omega_{1}^{\prime}}^{\prime},~b_{\omega_{2}^{\prime}%
}^{\prime\dag}\right]  =\left[  \tilde{b}_{\omega_{1}^{\prime}}^{\prime
},~\tilde{b}_{\omega_{2}^{\prime}}^{\prime\dag}\right]  =\delta\left(
\omega_{1}^{\prime}-\omega_{2}^{\prime}\right)  $. Thus the continuum limit
using regularization with the invariances modifies the UV entanglement
structure of typical states to generate a low-energy effective field theory.
Therefore, the maximal entanglement condition assumed by AMPS is not sustained
in this limit. This is the most fundamental reason why AMPS's firewalls do not
appear in the moving mirror model \cite{AS}. It seems very plausible that this
modification of the entanglement structure in the continuum limit may also
occur for the entanglement of $E$ and $L$ of the gravitational system to avoid
the original AMPS paradox.

The reason for no firewalls can be also understood from a viewpoint of
entanglement. In the AMPS arguement, it is assumed that quantum state of the
composite system is a typical pure state at a given time, in which a small
subsystem is almost maximally entangled with its complement subsystem. This
implies that entanglement entropy between them is proportional to volume of
the small subsystem. The entanglement history of black hole evaporation are
precisely imprinted into a sequence from early radiation to late radiation in
future null infinity region. Thus very early radiation should be almost
maximally entangled with other radiation. However, the radiation is described
by a low-energy field theory. Note that, in such a low-energy theory, the
entanglement entropy between a subsystem and its complement is proportional to
the boundary area, not the volume of the subsystem \cite{cirac}. Therefore, as
opposed to a naive expectation, the quantum state is not an almost maximally
entangled state. That is, the state is not a typical one. Actually it is
selected so as to satisfy natural conditions of the symmetries of field theory
and small excitation energy in the continuum limit. This fact clarifies that
the Page curve argument \cite{pagetime}, which is based on the typical state
assumption, is also incorrect as long as the description of low-energy field
theory is reproduced in null future infinity.

Next, we explain how the entropic paradox of firewalls related to Eq.
(\ref{1}) is resolved in the moving mirror model. First, the entanglement
entropy between the region $\left[  x_{1}^{-},x_{2}^{-}\right]  $ at the
future null infinity and its complement region is computed as:
\begin{equation}
S\left(  x_{1}^{-},x_{2}^{-}\right)  =\frac{1}{12}\ln\left(  \frac{\left(
f(x_{2}^{-})-f(x_{1}^{-})\right)  ^{2}}{\partial f(x_{2}^{-})\partial
f(x_{1}^{-})\epsilon_{1}^{-}\epsilon_{2}^{-}}\right)  , \label{11}%
\end{equation}
where $\epsilon_{1}^{-}$ ($\epsilon_{2}^{-}$) is a width cutoff of the
boundary at $x^{-}=x_{1}^{-}$ ($x^{-}=x_{2}^{-}$) \cite{hlw}. The outline of
derivation \ of Eq. (\ref{11}) is given in Appendix 1. Now, let us consider
$V$ as $E$ and a system composed of $AB$ and $\tilde{C}$ as $L$. It should be
stressed here that there is no positive-energy radiation in $E$. Thus, in this
case, $L$ is entangled with this fluctuation with zero energy. The idea that
quantum fluctuations with zero energy is entangled with Hawking radiation was
stressed first by Wilczek \cite{w} in the context of the information-loss problem.

Note that the late-part radiation $B$ in the original scenario of
\cite{firewall} corresponds to $B$ in Fig.~\ref{fig5}. Mode $C$ inside the
horizon of \cite{firewall} corresponds to $C$ in Fig.~\ref{fig2}, and so to
$\tilde{C}$ in Fig.~\ref{fig5}. In this model, the composite system of $B$ and
$C$ is fully entangled with a system composed of $A$, and the vacuum
fluctuation $V$ with zero energy. By using Eq. (\ref{11}), it turns out that
almost all of the entanglement of the $BC$ composite system is shared by $V$,
and the contribution of $A$ is negligibly small. This distribution of
entanglement can be attributed to the region width of $A^{\prime}B^{\prime}$
being much smaller than that of $V^{\prime}$ owing to the blueshift factor of
$A^{\prime}B^{\prime}$ at the past null infinity. Thus, the existence of this
entanglement shared by $V$ results in $S_{BC}>0$, and so the contradiction to
the strong subadditivity in Eq. (\ref{1}) is avoided. This observation
strongly suggests that entanglement between Hawking radiation and vacuum
fluctuation with zero energy also plays a crucial role in avoiding the
original entropic paradox of \cite{firewall} in quantum gravity.

What is wrong with the entropic argument of \cite{firewall} and \cite{mathur}
in this model? The authors of \cite{firewall} and \cite{mathur} assume that
the outside particle $B$ and the inside particle $C$ are in a purely entangled
state as
\[
\sum_{n}e^{-n\frac{\pi\omega}{a}}|n(\omega)\rangle_{B}|n(\omega)\rangle_{C}.
\]
This is a correct statement. However, mode $B$ has a long tail, which does not
vanish in the region of early-part radiation mode $A$. Thus, the separation
between systems $A$ and $B$ is not sufficient. This flaw does not change even
if we consider wave-packet modes by superposing one-particle states
\cite{h,takagi,AS2}. The wave packets indeed become localized to some extent;
however, they still have a long tail that merely decays through a power law.
\ Besides, if a localized wave packet basis is adopted, $|0_{in}\rangle$
cannot be exactly written in the form of Schmidt decomposition like Unruh
representation. The Schmidt decompostion of $|0_{in}\rangle$ is attained only
when the plane-wave mode functions are adopted for $ABV$. Thus, the locality
of $A$ and $B$ is not established completely. This implies that $B$ includes a
part of $A$. In such a situation, the strong subadditivity relation need not
hold true. In general, the strict localization of quantum states cannot be
attained by superposing one-particle states in relativistic quantum field
theory \cite{knight}. If we adopt strictly localized states as $A,B,$ and $C$,
the purity of the final state can be recovered only when we consider the whole
quantum system including the local vacuum part$~V$, as discussed above.

Besides, as mentioned above, $A~$and $B$ cannot share the (almost) maximum
entanglement as opposed to a native expectation from the discrete-model
analysis. In order to reproduce low-energy field theory with translational and
scale invariances, the high-energy entanglement structure of Eq. (\ref{e3})
should be modified as that of Eq. (\ref{06}). Thus $S_{\bar{B}_{A}B}=0$ and
$S_{\bar{B}_{A}BC}=S_{C}$ cannot be derived. Thus $I_{BC}=0$ is not correct.
Thus the entanglement structure in Eq. (\ref{06}) does not contradict the
no-drama condition with $I_{BC}\gg0$. In conclusion, no informational paradox
arises in the moving mirror model.

The extension of our result for two-dimensional moving mirrors to
four-dimensional gravitational collpase is not straightfoward. In this
analysis, we just consider its S mode contribution neglecting local curvature
effect. In order to treat the four dimensional case, we take account of higher
modes and potential terms induced by local curvature outside the horizon. It
is possible in principle that entanglement entropy contribution of higher
modes is evaluated using the two-dimensional models, but the formula in Eq.
(\ref{11}) cannot be applied because it is derived by use of conformal
symmetry. Besides, the height of the potential becomes larger in the final
stage of black hoke evaporation and may drastically change the fate of the
black hole.

Here, a comment is given about the renormalized entropy:
\begin{equation}
S_{ren}\left(  x_{1}^{-},x_{2}^{-}\right)  =\frac{1}{12}\ln\left(
\frac{\left(  f_{h}(x_{2}^{-})-f_{h}(x_{1}^{-})\right)  ^{2}}{\partial
f_{h}(x_{2}^{-})\partial f_{h}(x_{1}^{-})\left(  x_{2}^{-}-x_{1}^{-}\right)
^{2}}\right)  , \label{011}%
\end{equation}
which is provided in \cite{hlw}. Since $S_{ren}\left(  x_{1}^{-},x_{2}%
^{-}\right)  $ is defined as an excess entanglement with reference to the
vacuum entanglement \cite{AS2}, one might expect that $S_{ren}\left(
x_{1}^{-},x_{2}^{-}\right)  $ describes the entanglement between low-energy
excitations of the field by terminating the vacuum contribution. However, this
is not correct. In fact, $S_{ren}\left(  x_{1}^{-},x_{2}^{-}\right)  $ still
explicitly depends on the size of the local vacuum region. In fact, even when
the point $x^{-}=x_{1}^{-}$ belongs to the local vacuum region $V$ with
$f(x_{1}^{-})=x_{1}^{-}$ and $\partial f(x_{1}^{-})=1$, the renormalized
entropy still depends on $x_{1}^{-}$ as%

\[
S_{ren}\left(  x_{1}^{-},x_{2}^{-}\right)  =\frac{1}{12}\ln\left(
\frac{\left(  f_{h}(x_{2}^{-})-x_{1}^{-}\right)  ^{2}}{\partial f_{h}%
(x_{2}^{-})\left(  x_{2}^{-}-x_{1}^{-}\right)  ^{2}}\right)  .
\]
Thus, it is concluded that $S_{ren}\left(  x_{1}^{-},x_{2}^{-}\right)  $ still
includes the vacuum entanglement contribution. Note that the enormous amount
of vacuum state entanglement, which is guaranteed by the Reeh--Schlieder
theorem, includes a high-energy contribution as well as a low-energy
contribution, because an arbitrary state of $L$ can be generated independent
of its excitation energy using a local operator acting on $E$. No natural
threshold separating the high-energy contribution from the low-energy
contribution of $S$ is known in information theory, because the entanglement
itself is a purely informational concept independent of energy.\ In fact, when
self-interactions of the field exist, entanglement mixing between low-energy
modes and high-energy modes occurs. Even the finiteness of $S_{ren}\left(
x_{1}^{-},x_{2}^{-}\right)  $ for general quantum states has not yet been
proven. It is also unclear that the vanishing value of $S_{ren}\left(
x_{1}^{-},x_{2}^{-}\right)  $ implies the purity of the state of local
excitations. Besides, it should be stress that $S_{ren}\left(  x_{1}^{-}%
,x_{2}^{-}\right)  $ does not satisfy the strong additivity as shown in
Appendix 2, though it exactly holds for $S\left(  x_{1}^{-},x_{2}^{-}\right)
$. At present, the meaning of $S_{ren}\left(  x_{1}^{-},x_{2}^{-}\right)  $
remains elusive in the context of quantum information theory. Meanwhile,
$S\left(  x_{1}^{-},x_{2}^{-}\right)  $ in Eq. (\ref{11}) is an appropriate
measure of the entanglement to treat high-energy entanglement and low-energy
entanglement simultaneously and systematically.

\section{Emergence and Resolution of Firewall Measurement Paradox}

\bigskip

~

It is assumed that the above firewall prohibition is because we do not
actually perform any measurements of $E$. In contrast with AMPS's original
scenario, if we really perform a measurement of $E$ to gain the information of
$L$, another version of the firewall paradox arises in this model. A similar
paradox appears in a more simple model of Rindler horizon\cite{ref}. The
Reeh--Schlieder theorem \cite{rs} asserts that the set of states generated
from $|0_{in}\rangle$ through the polynomial algebra of local operators in any
bounded spacetime region is dense in the entire Hilbert space of the field.
Thus, in principle, any state of $L^{\prime}$ can be arbitrarily reproduced
closely by operating a polynomial of local $E^{\prime}$ operators on
$|0_{in}\rangle~$\cite{PT}. This is the same for $E$ and $L$. Assuming
finiteness of the Hilbert space dimensions, a measurement operator \cite{nc}
of $E$, the post-measurement state of which includes the firewall, can be
constructed. As a result, the observer encounters a firewall as explained below.

\begin{figure}[t]
\begin{center}
\includegraphics[height=.5\textheight]{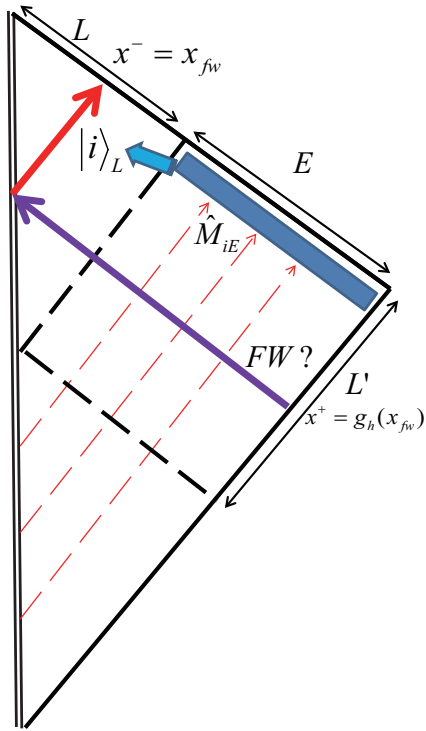}
\end{center}
\caption{(color online). Penrose diagram representing the measurement of $E$
to obtain the information of $L$. The red and violet arrows denote the
firewall.}%
\label{fig8}%
\end{figure}

\begin{figure}[t]
\begin{center}
\includegraphics[height=.5\textheight]{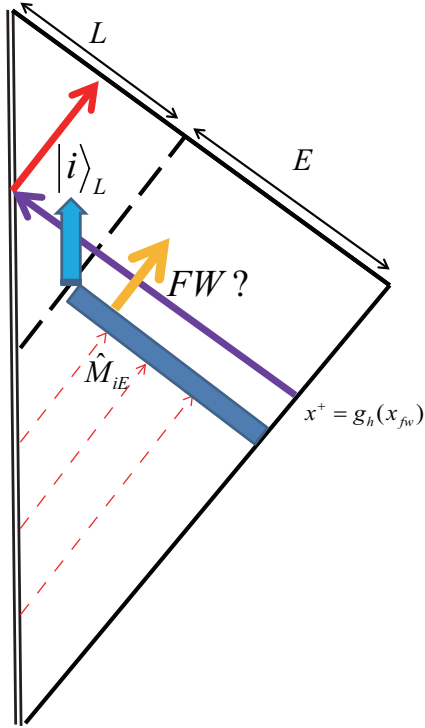}
\end{center}
\caption{(color online). Penrose diagram representing the firewall paradox in
this model. An observer in motion like the yellow arrow will encounter the
firewall.}%
\label{fig9}%
\end{figure}

\begin{figure}[t]
\begin{center}
\includegraphics[height=.4\textheight]{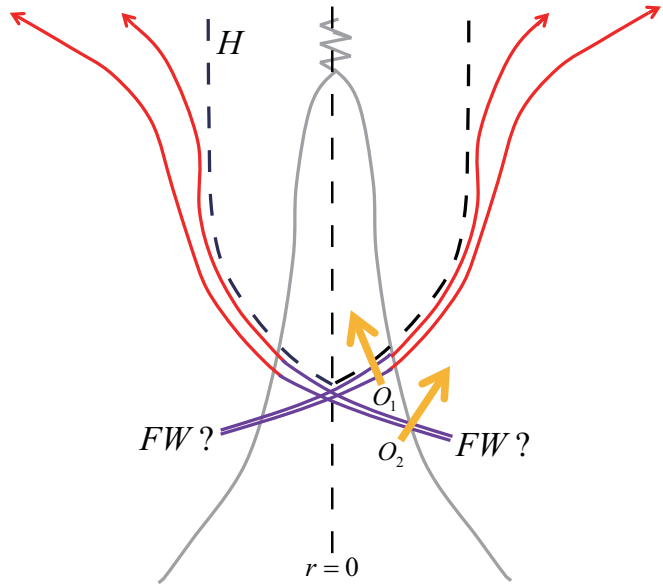}
\end{center}
\caption{(color online). Schematic diagram of gravitational collapse with
firewalls. Observer $O_{1}$ encounters an outgoing firewall in the original
paradox of AMPS. Observer $O_{2}$ encounters an incoming firewall, as well as
in the moving mirror model.}%
\label{fig10}%
\end{figure}

First, it is worth recalling that $|0_{in}\rangle$ is translationally
invariant: $x^{+}\rightarrow x^{+}+c^{+}$. Thus, the entanglement between
$L^{\prime}$ and $E^{\prime}$ is independent of the boundary position. Thus,
the entanglement is also independent of where the $L$--$E$ boundary is fixed
at the future null infinity.\ Let us consider a general measurement of quantum
information theory \cite{nc} for the vacuum fluctuation of $E$ that outputs
the result $i$ \cite{QET}. Based on the Reeh--Schlieder theorem, let us
consider that, besides the background Hawking radiation, a wave packet with
positive energy of the order of the radiation temperature appears at
$x^{-}=x_{fw}$ in the post-measurement state $|i\rangle_{L}$ of $L$. As
depicted in Fig.~\ref{fig8}, the wave packet traveling back in time is
severely blueshifted, and it carries a huge amount of energy in $L^{\prime}$
at the past null infinity. The deadly energetic flux emerges at $x^{+}%
=g_{h}(x_{fw})$, where $g_{h}(x)$ is a solution of $f_{h}\left(
g_{h}(x)\right)  =x$. Therefore, if the measurement is performed before the
firewall forms, as depicted in Fig.~\ref{fig9}, an observer attempting to move
along the arrow will burn out by the firewall. This is the firewall
measurement paradox. \ If the measurement is not executed, the observer safely
passes over $x^{+}=g_{h}(x_{fw})$, in contrast to AMPS's original scenario.\ A
similar version of the firewall measurement paradox arises in the original
black hole scenario as well. Figure~\ref{fig10} schematically depicts a
gravitational collapse in which the outgoing firewall attacks an infalling
observer $O_{1}$, as argued in \cite{firewall}. The second firewall discussed
above appears in the incoming modes and will attack a different observer
$O_{2}$ in Fig.~\ref{fig10}. The paradox is similar to the famous Unruh-Wald
argument that a particle detector with uniform acceleration generates wave
packets in the post-measurement states in a causally disconnected region
\cite{UW}. However, it should be stressed that general measurements should be
analyzed in the firewall arguments, as opposed to the Unruh-Wald one. In the
next section, we argue that the firewall does not appear, provided the
measurement energy cost is much smaller than the ultraviolet cutoff scale.
This leads to the firewall information censorship conjecture.

In the moving mirror models, the time evolution is simple. The plane-wave
modes $\exp(-i\omega x^{+})$ of the free field are just stretched into
$\exp(-i\omega f_{h}(x^{-}))$ by the mirror trajectory without loss of
unitarity and entanglement. Thus, a late-time measurement of $E$ can be
replaced by an early-time measurement of $E^{\prime}$, which yields the same
results and post-measurement states. To understand this useful fact more
concretely, let us consider a simple example of a general measurement
\cite{nc} of $E$ outputting results $\mu$ with measurement operators%
\[
\hat{M}_{\mu E}=F_{\mu}\left(  \int\Omega(x^{-})\partial_{x^{-}}\hat{\varphi
}_{out}\left(  x^{-}\right)  dx^{-}\right)  ,
\]
where $\Omega(x^{-})~$is a real window function for $E$. The function $F_{\mu
}\left(  x\right)  $ must satisfy $\sum_{\mu}\left\vert F_{\mu}\left(
x\right)  \right\vert ^{2}=1$ to impose the unitarity condition on the
measurement operators \cite{nc}. Owing to the relation $\hat{\varphi}%
_{out}\left(  x^{-}\right)  =\hat{\varphi}_{in}\left(  f_{h}(x^{-})\right)  $,
this measurement corresponds to a general measurement of $E^{\prime}$ with
measurement operators defined as%

\[
\hat{M}_{\mu E^{\prime}}=F_{\mu}\left(  \int\Omega\left(  g_{h}(x^{+})\right)
\partial_{x^{+}}\hat{\varphi}_{in}\left(  x^{+}\right)  dx^{+}\right)  .
\]

\begin{figure}[t]
\begin{center}
\includegraphics[height=.5\textheight]{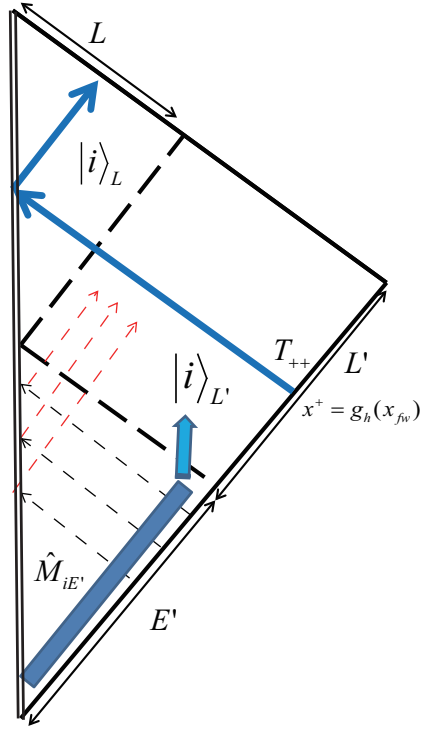}
\end{center}
\caption{(color online). Penrose diagram representing an early-time
measurement of $E^{\prime}$, which provides the same influence as the
late-time measurement in Fig.~\ref{fig9}.}%
\label{fig11}%
\end{figure}

Notice that $\Omega\left(  g_{h}(x^{+})\right)  $ indeed becomes a window
function for $E^{\prime}$ at the past null infinity. Taking account of such a
correspondence, we consider an arbitrary measurement of $E^{\prime}$
outputting $i$ with measurement operator $\hat{M}_{iE^{\prime}}$ obeying the
unitary condition%
\begin{equation}
\sum_{i}\hat{M}_{iE^{\prime}}^{\dag}\hat{M}_{iE^{\prime}}=\hat{I}. \label{20}%
\end{equation}
The measurement is depicted in Fig.~\ref{fig11}. Our task is to check whether
the firewall exists in the post-measurement state $\hat{M}_{iE^{\prime}%
}|0_{in}\rangle$.

When $x^{+}$ takes values around $g_{h}(x_{fw})$ of $L^{\prime}$, the
expectation value of the energy flux for a fixed $i$ is computed as%
\begin{equation}
\left\langle T_{++}(x^{+})\right\rangle _{i}\varpropto\langle0_{in}|\hat
{M}_{iE^{\prime}}^{\dag}\hat{T}_{++}(x^{+})\hat{M}_{iE^{\prime}}|0_{in}%
\rangle=\langle0_{in}|\hat{\Pi}_{iE^{\prime}}\hat{T}_{++}(x^{+})|0_{in}%
\rangle, \label{7}%
\end{equation}
where $\hat{\Pi}_{iE^{\prime}}=\hat{M}_{iE^{\prime}}^{\dag}\hat{M}%
_{iE^{\prime}}$ is a positive operator valued measure (POVM) of the
measurement. Note that $\left\langle T_{++}(x^{+})\right\rangle _{i}$ is just
a two-point correlation function of $\hat{\Pi}_{iE^{\prime}}^{\prime}$ and
$\hat{T}_{++}(x^{+})$ in the vacuum state. The correlation functions for
nonsingular $\hat{\Pi}_{iE^{\prime}}$ simply decay via a power law as a
function of the distance $l$ between the boundary of $E^{\prime}$ ($x^{+}%
\sim-1/\kappa$) and the would-be firewall position ($x^{+}=g_{h}(x_{fw})$).
Therefore, an arbitrary nonsingular measurement is incapable of generating any
outstanding peak of $\left\langle T_{++}(x^{+})\right\rangle _{i}$ at
$x^{+}=g_{h}(x_{fw})$. This implies no firewall for any $i$.

For example, in a one-bit measurement of $E^{\prime}$ outputting $i=0,1$ with%
\begin{align*}
\hat{M}_{0E^{\prime}}  &  =\cos\left(  \int_{-\infty}^{x_{E^{\prime}}}%
\lambda_{E^{\prime}}(x^{+})\partial_{x^{+}}\hat{\varphi}_{in}\left(
x^{+}\right)  dx^{+}\right)  ,\\
\hat{M}_{1E^{\prime}}  &  =\sin\left(  \int_{-\infty}^{x_{E^{\prime}}}%
\lambda_{E^{\prime}}(x^{+})\partial_{x^{+}}\hat{\varphi}_{in}\left(
x^{+}\right)  dx^{+}\right)  ,
\end{align*}
the correlation functions for $x^{+}>x_{E^{\prime}}$ are computed as%
\[
\langle0|\hat{\Pi}_{iE^{\prime}}\hat{T}_{++}(x^{+})|0\rangle=2\left(
-1\right)  ^{i+1}\left\vert \int_{-\infty}^{x_{E^{\prime}}}G(x^{+}-x^{\prime
+})\lambda_{E^{\prime}}(x^{\prime+})dx^{\prime+}\right\vert ^{2},
\]
where $\lambda_{E^{\prime}}(x^{+})$ is a real function that vanishes at the
outside region of $E^{\prime}$,
\[
G(x^{+}-x^{\prime+})=\langle0_{in}|\partial_{x^{+}}\hat{\varphi}_{in}\left(
x^{+}\right)  \partial_{x^{+}}\hat{\varphi}_{in}\left(  x^{\prime+}\right)
|0_{in}\rangle=-\frac{1}{4\pi\left(  x^{+}-x^{\prime+}-i0\right)  ^{2}},
\]
and it is assumed that $E^{\prime}$ lies in the region $(-\infty,x_{E^{\prime
}})$. Thus, for both $i$, no peak of energy density appears at any $x^{+}$ of
$L^{\prime}$.

\begin{figure}[t]
\begin{center}
\includegraphics[width=.9\textwidth]{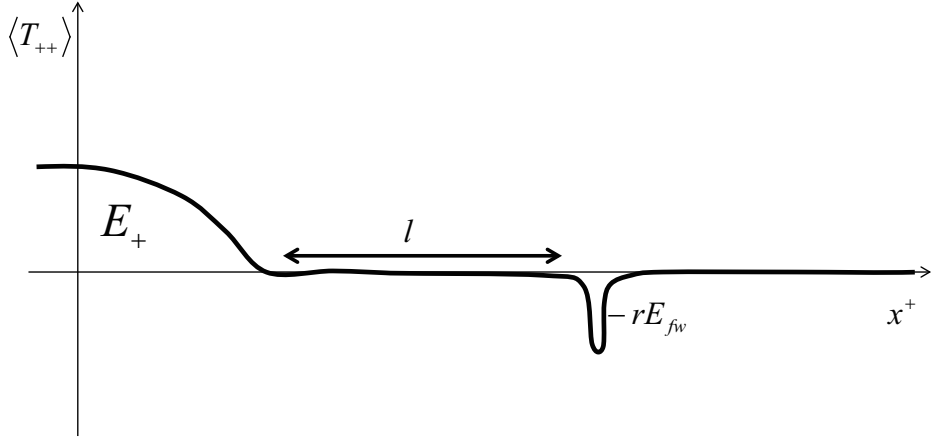}
\end{center}
\caption{Schematic diagram representing the average energy flux distribution
in $\hat{M}_{jE^{\prime}}|0_{in}\rangle$.}%
\label{fig12}%
\end{figure}

If one wants to create a post-measurement state $|i\rangle_{L}$ involving the
firewall, quite singular measurement operators should be invoked. Precisely,
the total energy after the measurement,%
\[
\left\langle \hat{H}\right\rangle =\sum_{i}\int_{-\infty}^{\infty}%
\langle0_{in}|\hat{M}_{iE^{\prime}}^{\dag}\hat{T}_{++}(x^{+})\hat
{M}_{iE^{\prime}}|0_{in}\rangle dx^{+},
\]
must be divergent. For instance, let us imagine a post-measurement state
$\hat{M}_{iE^{\prime}}|0_{in}\rangle$ for a fixed $i$ with a sharp energy peak
at $x^{+}=g_{h}(x_{fw})$, the energy flux of which is approximated by
\begin{equation}
\frac{\langle0_{in}|\hat{\Pi}_{iE^{\prime}}\hat{T}_{++}(x^{+})|0_{in}\rangle
}{\langle0_{in}|\hat{\Pi}_{iE^{\prime}}|0_{in}\rangle}=E_{fw}\delta\left(
x^{+}-g_{h}(x_{fw})\right)  \label{12}%
\end{equation}
for $x^{+}>x_{E^{\prime}}$. Here, $E_{fw}$ is the energy of the firewall. From
the unitarity condition of the measurement operators in Eq. (\ref{20}), the
total sum of contributions for the measurement results vanishes:%
\[
\sum_{j}\langle0_{in}|\hat{\Pi}_{jE^{\prime}}\hat{T}_{++}(x^{+})|0_{in}%
\rangle=\langle0_{in}|\hat{T}_{++}(x^{+})|0_{in}\rangle=0.
\]
Hence, without loss of generality, we can assume that there is an output $j$
that satisfies%

\[
\frac{\langle0_{in}|\hat{\Pi}_{jE^{\prime}}\hat{T}_{++}(x^{+})|0_{in}\rangle
}{\langle0_{in}|\hat{\Pi}_{jE^{\prime}}|0_{in}\rangle}=-rE_{fw}\delta\left(
x^{+}-g_{h}(x_{fw})\right)  ,
\]
where $x^{+}>x_{E^{\prime}}$ and $r=\langle0_{in}|\hat{\Pi}_{iE^{\prime}%
}|0_{in}\rangle/\langle0_{in}|\hat{\Pi}_{jE^{\prime}}|0_{in}\rangle>0$. In
fact, this condition holds for a general measurement with one-bit measurement
operators such that
\[
\hat{M}_{0}^{\prime}=\hat{M}_{iE^{\prime}},~\hat{M}_{1}^{\prime}=\sqrt
{I-\hat{M}_{iE^{\prime}}^{\dag}\hat{M}_{iE^{\prime}}}.
\]
Because $|0_{in}\rangle$ is a fully entangled state, $|0_{in}\rangle$ is an
eigenstate of neither $\hat{\Pi}_{iE^{\prime}}$ nor $\hat{\Pi}_{jE^{\prime}}$,
and $r$ is naturally anticipated as $O(1)$. This signifies that we have a
negative-energy shock wave at $x^{+}=g_{h}(x_{fw})$. Let $l$ denote the
distance between the shock wave and $E^{\prime}$:%
\[
l=g_{h}(x_{fw})-x_{E^{\prime}}.
\]
To maintain the nonnegativity of the total Hamiltonian of the field in
$\hat{M}_{jE^{\prime}}|0_{in}\rangle$, there exists a positive-energy region
of $E^{\prime}$. The energy flux distribution is schematically depicted in
Fig.~\ref{fig12}. A general bound of negative energy for any quantum state in
which the total energy is finite \cite{hmy} enables us to derive the following
bound on the firewall energy:
\begin{equation}
E_{fw}<\frac{1}{12\pi rl}. \label{9}%
\end{equation}
This clearly contradicts the emergence of a firewall with a huge energy flux
of the order of the ultraviolet cutoff scale. Thus, actually, a firewall with
deadly high energy is not present, as long as the total energy of the
post-measurement state is small. The derivation of the bound in Eq. (\ref{9})
is outlined as follows.

Let us consider a nonnegative continuous function $\xi(x)$ and define a
Hermitian operator as%

\[
\hat{H}_{\xi}=\int_{-\infty}^{\infty}\xi(x^{+})\hat{T}_{++}(x^{+})dx^{+}.
\]
Then, the inequality
\[
\operatorname*{Tr}\left[  \hat{H}_{\xi}\hat{\rho}\right]  \geq-\frac{1}{12\pi
}\int_{-\infty}^{\infty}\left(  \partial_{x}\sqrt{\xi(x)}\right)  ^{2}dx
\]
holds for an arbitrary state $\hat{\rho}~$\cite{F}. Now consider a state
$\hat{\rho}$ satisfying
\[
\operatorname*{Tr}\left[  \hat{T}_{++}(x^{+})\hat{\rho}\right]  =-rE_{fw}%
\delta\left(  x^{+}-g_{h}(x_{fw})\right)
\]
for $x^{+}>x_{E^{\prime}}$. Let us impose the values $\xi(x)=0$ for
$x\in(-\infty,x_{E^{\prime}})$ and $\xi(x)=1$ for $x\in\left(  g_{h}%
(x_{fw})-0,\infty\right)  $ on $\xi(x)$. As a result, $\operatorname*{Tr}%
\left[  \hat{H}_{\xi}\hat{\rho}\right]  =-rE_{fw}$ for an arbitrary $\xi(x)$
satisfying the above conditions. Thus,
\[
rE_{fw}\leq\frac{1}{12\pi}\inf_{\xi(x)}\int_{-\infty}^{\infty}\left(
\partial_{x}\sqrt{\xi(x)}\right)  ^{2}dx
\]
must be satisfied. The infimum of the $\xi(x)$ satisfying the above boundary
conditions is then taken. By using a variation method, the infimum is obtained
from a function $\xi_{opt}(x)$ obeying
\[
\xi_{opt}(x)=\left(  \frac{x-x_{E^{\prime}}}{l}\right)  ^{2}%
\]
for $x\in(x_{E^{\prime}},g_{h}(x_{fw}))$, so the inequality of Eq. (\ref{9})
can be derived. When $E_{fw}$ approaches $\frac{1}{12\pi rl}$, the positive
energy $E_{+}$ of $E^{\prime}$, which is injected by the measurement device,
is considered to diverge to infinity. In fact, when $\hat{M}_{jE^{\prime}%
}|0_{in}\rangle$ is a squeezed state for example, it can be explicitly proven
that $E_{+}$ obeys the following inequality:
\begin{equation}
E_{+}\geq\frac{rE_{fw}}{1-12\pi lrE_{fw}}, \label{8}%
\end{equation}
where the right-hand side diverges to infinity in the limit of $E_{fw}%
\rightarrow\frac{1}{12\pi rl}$. The proof of Eq. (\ref{8}) is provided in the
appendix 3. Recall that $E$ is in a local vacuum state with zero energy
because the mirror does not move yet at the reflection time of the modes.
Hence, when the measurement is performed for $E$ as in Fig.~\ref{fig9}, the
measurement energy is the same as that for $E^{\prime}$, and we are able to
conclude that no firewall appears provided the measurement energy of $E$ is
much smaller than the cutoff scale.

A similar paradox may be considered in a Rindler horizon model \cite{ref}.
Imagine an observer who measures the Rindler energy, thereby projecting the
state onto a Rindler energy-eigenstate. It is easy to check that in this
state, the two point function of a scalar field across the Rindler horizon is
ill-behaved. This is the analogue of a "firewall." Note that local
measurements generally inject energy on average to the field in $|0_{in}%
\rangle$ owing to the passivity of the state \cite{passivity}. Hence, quantum
measurements always require an energy cost. Though the Reeh--Schlieder theorem
is mathematically correct, it does not guarantee that the measurement energy
required to create $|i\rangle_{L}$ is finite even if the measurement operator
exists. When we measure Rindler particle number eigenstates $|n(\omega
)\rangle_{E^{\prime}}$ of Unruh modes in Eq. (\ref{6}) to create
$|n(\omega)\rangle_{L^{\prime}}$, energy of the order of the ultraviolet
cutoff scale is injected at the boundary of $E^{\prime}$ and $L^{\prime}$.
This huge amount of energy diffuses along the future Rindler horizons and will
modify spacetime drastically. Beyond our model, singular measurements
generally require the preparation of a divergent amount of energy in the
measurement region before the measurement is performed, and this energy is
expected to provide a large back reaction to spacetime. The effect may cause
the formation of a new black hole in the measurement region and enclose the
measurement device within the event horizon before it outputs results.
Therefore, the estimation of measurement energy is quite important in any
thought experiment on firewalls with measurements. This is the main message of
this paper. Based on this consideration, we propose a conjecture of firewall
information censorship whereby the leakage of the measurement information on
encounters with firewalls is prohibited at a profound level.

Before closing this section, we need to comment on measurements of thermal
radiation. In the above argument, the measurement is performed in the local
vacuum region. However, in the original scenario of \cite{firewall}, the
measurement is considered in the early-part radiation region to predict the
behavior of the late-part radiation. When we consider the same setup in the
moving mirror model, no firewall paradox has a definite meaning because the
available measurement region is too small to allow an observer to perform the
measurement before encountering firewalls. However, if one wants to infer the
existence of firewalls in a past region, it is interesting to treat the early
Hawking-like radiation as $E$. Let us consider a nonsingular general
measurement of $E$ with measurement operators that output $\mu$:
\[
\hat{M}_{\mu E}=F_{\mu}\left(  \int\Omega_{E}\left(  x^{-}\right)
\partial_{x^{-}}\hat{\varphi}_{out}\left(  x^{-}\right)  dx^{-}\right)  .
\]
When the result $\mu$ of $E$ is obtained, let us introduce an
energy--momentum-tensor gain of $L$ as
\[
\Delta^{(\mu)}T_{--}(x_{L}^{-})=\frac{\langle0_{in}|\hat{M}_{\mu E}^{\dag}%
\hat{M}_{\mu E}T_{--}(x_{L}^{-})|0_{in}\rangle}{\langle0_{in}|\hat{M}_{\mu
E}^{\dag}\hat{M}_{\mu E}|0_{in}\rangle}-\langle0_{in}|T_{--}(x_{L}^{-}%
)|0_{in}\rangle.
\]
Note that the Hawking-like radiation regime is well approximated by the moving
mirror trajectory in Eq. (\ref{3}). By using Eq. (\ref{3}) and $x_{L}%
^{+}=f_{h}(x_{L}^{-})$, the gain of $T_{++}$ at the near horizon ($x_{L}%
^{+}\sim0$) is evaluated as%

\[
\Delta^{(\mu)}T_{++}(x_{L}^{+})=\left(  \frac{\partial x_{L}^{-}}{\partial
x_{L}^{+}}\right)  ^{2}\Delta^{(\mu)}T_{--}(x_{L}^{-})\sim\frac{1}{\left(
\kappa x_{L}^{+}\right)  ^{2}}\Delta^{(\mu)}T_{--}(x_{L}^{-}).
\]
If $\Delta^{(\mu)}T_{--}(x_{L}^{-})$ does not vanish when $x_{L}%
^{-}\rightarrow\infty$ ($x_{L}^{+}\rightarrow0$), we confirm the existence of
a firewall. However, in the thermal region, the two-point correlation function
of $\partial_{x^{-}}\hat{\varphi}_{out}\left(  x^{-}\right)  $ behaves as%
\begin{align*}
\langle0_{in}|\partial_{x^{-}}\hat{\varphi}_{out}\left(  x_{L}^{-}\right)
\partial_{x^{-}}\hat{\varphi}_{out}\left(  x_{E}^{-}\right)  |0_{in}\rangle &
=-\frac{\kappa^{2}}{16\pi\sinh^{2}\left(  \frac{\kappa}{2}\left(  x_{L}%
^{-}-x_{E}^{-}-i0\right)  \right)  }\\
&  \sim O\left(  \exp\left(  -\kappa x_{L}^{-}\right)  \right)  .
\end{align*}
Using this correlation, it is possible to show for nonsingular measurements
that $\Delta^{(\mu)}T_{--}(x_{L}^{-})=O(\exp\left(  -2\kappa x_{L}^{-}\right)
)~=O(\left(  \kappa x_{L}^{+}\right)  ^{2})~$ when considering $x_{L}^{-}%
\sim\infty$. This implies that $\Delta^{(\mu)}T_{++}(x_{L}^{+})$ is always
finite even if we consider $x_{L}^{+}\rightarrow0$. Hence, no firewall appears
in this case either, as long as the measurement is not singular.

\bigskip

\section{Summary}

\bigskip

~

In a moving mirror model with a free massless scalar field, no AMPS's
firewalls appear even though the final state is pure and fully entangled. The
reason is that the continuum limit to a low-energy field theory with conformal
invariance modifies the UV entanglement structure of almost maximally
entangled states of the discretized systems. The entropic paradox does not
arise in this model. The $BC$ system in this model is fully entangled with the
early vacuum fluctuation $V$. This implies that $S_{BC}>0$ and allows us to
avoid violation of the strong subadditivity in Eq. (\ref{1}). We have examined
an extended firewall paradox with quantum measurement of the early modes $E$
with the trajectory in Eq. (\ref{10}). The dynamics of the model is quite
simple and exactly unitary. The in-vacuum state $|0_{in}\rangle$ provides
enough entanglement between $E$ and the late modes $L$. It is proven that, as
long as nonsingular measurements are adopted, deadly energetic firewalls do
not emerge in this case. The crucial point is the energy cost of quantum
measurements of $E$. Generation of firewalls via entanglement between $E~$and
$L$ requires\ quite singular measurements, the energy cost of which is of the
order of the ultraviolet cutoff scale, i.e., the Planck scale. Preparation of
such a huge measurement energy in a region of $E$ may provide large back
reaction to spacetime before the measurement is executed and may change the
problem itself drastically. This consideration leads to the firewall
information censorship conjecture, whereby the measurement information leakage
of encounters with firewalls is prohibited at a profound level.

\bigskip

\textbf{Acknowledgments}\newline

We would like to thank Bill Unruh and Joe Polchinski for related discussions.
Also we would like to thank one of reviewers in Physical Review D for pointing
out the simple model of Rindler horizon using the Reeh--Schlieder theorem. M.
H. would like to thank the organizers of the KIAS-YITP joint workshop,
\textquotedblleft String Theory, Black Holes and Holography" for their hospitality.

\bigskip{\LARGE Appendix 1: Proof Outline of HLW Formula}

In this appendix, an outline of derivation of Eq. (\ref{11}) in \cite{hlw} is
provided by using conformal symmetry. Let us think a massless scalar field
$\hat{\varphi}\left(  x\right)  $ in one dimension. Consider a Rindler
observer with acceleration $\kappa$. The corresponding Rindler coordinates are
given by%

\begin{align*}
t  &  =\frac{1}{\kappa}e^{\kappa u}\sinh(\kappa\tau),\\
x  &  =\frac{1}{\kappa}e^{\kappa u}\cosh(\kappa\tau),
\end{align*}
where the trajectory of the observer is given by $u=0$. The field is observed
in a thermal state at temperature $T=\frac{\kappa}{2\pi}$ due to Unruh effect.
Because the entropy density in one dimension is computed as $\varrho=\frac{\pi
T}{3}$, total entanglement entropy between $\left[  u_{1},u_{2}\right]  $ and
its complement is evaluated as
\[
S=\varrho(u_{2}-u_{1}).
\]
In the original coordiate $x$, this can be written as%
\[
S=\frac{\varrho}{\kappa}\ln\frac{x_{2}}{x_{1}}=\frac{1}{6}\ln\frac{x_{2}%
}{x_{1}}.
\]
By setting \thinspace$x_{1}=\epsilon_{o}\sim0$ and $x_{2}=L_{o}\sim\infty$, we
obtain%
\[
S=\frac{1}{6}\ln\frac{L_{o}}{\epsilon_{o}}.
\]
Here let us introduce new coordinates $\sigma^{\pm}$ as%
\[
t\pm x=\frac{\sin\left(  \frac{\pi}{L}(l-\sigma^{\pm})\right)  }{\sin\left(
\frac{\pi}{L}\sigma^{\pm}\right)  },
\]
where $l$ is a positive parameter satisfing $0<l<L$ and $L$ is circumference
of the mapped region, which is a circle. The $S$ describes entanglement
entropy between $\left[  0,l\right]  $ and $\left[  l,L\right]  $. By taking
$L\rightarrow\infty$ limit, we obtain entanglement entropy of the field in
open space $\left(  -\infty,\infty\right)  $ between $\left[  0,l\right]  $
and its complement as%

\begin{equation}
S=\frac{1}{12}\ln\frac{l^{2}}{\epsilon_{1}\epsilon_{2}}. \label{1001}%
\end{equation}
Here $\epsilon_{1},\epsilon_{2}$ are ultraviolet cutoffs defined by
\begin{align*}
\epsilon_{o}  &  =\frac{\pi}{L}\frac{\epsilon_{2}}{\sin\left(  \frac{\pi l}%
{L}\right)  },\\
L_{o}  &  =\frac{L}{\pi\epsilon_{1}}\sin\left(  \frac{\pi l}{L}\right)  .
\end{align*}
Let us apply the formula in Eq. (\ref{1001}) to the in-vacuum state of the
moving mirror model.
\[
S=\frac{1}{12}\ln\left(  \frac{\left(  x_{2}^{+}-x_{1}^{+}\right)  ^{2}%
}{\epsilon_{1}^{+}\epsilon_{2}^{+}}\right)  .
\]
During the scattering by the mirror, entanglement is preserved. Thus we can
estimate $S$ in terms of the out states as%

\[
S\left(  x_{1}^{-},x_{2}^{-}\right)  =\frac{1}{12}\ln\left(  \frac{\left(
f(x_{2}^{-})-f(x_{1}^{-})\right)  ^{2}}{\partial f(x_{2}^{-})\partial
f(x_{1}^{-})\epsilon_{1}^{-}\epsilon_{2}^{-}}\right)  ,
\]
where $x_{a}^{+}=f(x_{a}^{-})$ and $\epsilon_{a}^{+}=\partial f(x_{a}%
^{-})\epsilon_{a}^{-}$ for $a=1,2$. This coincides with Eq. (\ref{11}).

\newpage

\newpage{\LARGE Appendix 2: Sudadditivy Breaking of Renormalized Entangement
Entropy}

Let us consider a monotonically increasing function $f(x)$ and three regions
$A,B,C$ given by%

\[
A=\left[  0,l\right]  ,B=\left[  l,2l\right]  ,C=\left[  2l,3l\right]  ,
\]
with positive $l$. Then the strong subadditivity imposes
\begin{equation}
\Delta S_{ren}=S_{ren}^{AB}+S_{ren}^{BC}-S_{ren}^{B}-S_{ren}^{ABC}\geq0,
\label{e1}%
\end{equation}
where $S_{ren}^{AB}=S_{ren}(0,2l),S_{ren}^{BC}=S_{ren}(l,3l),S_{ren}%
^{B}=S_{ren}(l,2l)~$\ and $S_{ren}^{ABC}=S_{ren}(0,3l)$. Since $\Delta
S_{ren}$ is computed as
\[
\Delta S_{ren}=\frac{1}{6}\ln\frac{3\left(  f(2l)-f(0)\right)  \left(
f(3l)-f(l)\right)  }{4\left(  f(2l)-f(l)\right)  \left(  f(3l)-f(0)\right)
},
\]
Eq. (\ref{e1}) means that
\[
\frac{\left(  f(2l)-f(0)\right)  \left(  f(3l)-f(l)\right)  }{\left(
f(2l)-f(l)\right)  \left(  f(3l)-f(0)\right)  }\geq\frac{4}{3}.
\]
However, when we assume $f(0)=0$ and $f(l)=\epsilon$ with infinitesimal
positive $\epsilon$, it is easily verified that
\[
\lim_{\epsilon\rightarrow+0}\frac{\left(  f(2l)-f(0)\right)  \left(
f(3l)-f(l)\right)  }{\left(  f(2l)-f(l)\right)  \left(  f(3l)-f(0)\right)
}=1<\frac{4}{3}.
\]
Hence the strong subadditivity of $S_{ren}\left(  x_{1}^{-},x_{2}^{-}\right)
$ is explicitly broken.

\bigskip

\newpage{\LARGE Appendix 3:\ Proof of Eq. (\ref{8})}

{\LARGE \bigskip}

In this appendix, we prove the bound in Eq. (\ref{8}). Let us assume that the
state $|\Psi\rangle\varpropto\hat{M}_{jE^{\prime}}|0_{in}\rangle$ is a
squeezed state defined by $\hat{c}_{\omega}|\Psi\rangle=0$ with annihilation
operators $\hat{c}_{\omega}$ satisfying $\left[  \hat{c}_{\omega},\hat
{c}_{\omega^{\prime}}^{\dag}\right]  =\delta\left(  \omega-\omega^{\prime
}\right)  $. By using the in-field $\hat{\varphi}_{in}$, $\hat{c}_{\omega}$ is
constructed as
\[
\hat{c}_{\omega}=\frac{i}{\sqrt{\pi\omega}}\int_{-\infty}^{\infty}\exp\left(
i\omega F(x^{+})\right)  \partial_{x^{+}}\hat{\varphi}_{in}\left(
x^{+}\right)  dx^{+},
\]
where $F(x)$ is a monotonically increasing function satisfying $F(\pm\infty)$
$=\pm\infty$. The average energy flux is computed as%

\[
\langle\Psi|\hat{T}_{++}(x^{+})|\Psi\rangle=-\frac{1}{24\pi}\left[
\frac{\partial_{x}^{3}F}{\partial_{x^{+}}F}-\frac{3}{2}\left(  \frac
{\partial_{x^{+}}^{2}F}{\partial_{x^{+}}F}\right)  ^{2}\right]  ,
\]
and the total energy is given by%
\[
E_{tot}=\int_{-\infty}^{\infty}\langle\Psi|T_{++}(x^{+})|\Psi\rangle
dx^{+}=\frac{1}{48\pi}\int_{-\infty}^{\infty}\left(  \frac{\partial_{x^{+}%
}^{2}F}{\partial_{x^{+}}F}\right)  ^{2}dx^{+}.
\]
By shifting the origin of $x^{+}$, we let the $E^{\prime}$ station be at
$x^{+}\leq0$ and the shock wave with negative energy $-rE_{fw}$ be at
$x^{+}=l$. The energy flux distribution for $x^{+}>0$ is given by
\begin{equation}
\langle\Psi|\hat{T}_{++}(x^{+})|\Psi\rangle=-rE_{fw}\delta\left(
x^{+}-l\right)  . \label{13}%
\end{equation}
For $x>0$, the most general form of $F\left(  x\right)  $ satisfying Eq. (13)
is solved\ as%

\begin{equation}
F\left(  x\right)  =\frac{a+b\left(  x-l\right)  }{c+d\left(  x-l\right)
}\Theta\left(  l-x\right)  +\left[  \frac{a}{c}-\frac{ad-bc}{c^{2}}\left(
x-l\right)  \right]  \Theta\left(  x-l\right)  , \label{15}%
\end{equation}
where $\Theta\left(  x\right)  $ is the Heaviside step function, and $a$, $b$,
$c$, and $d$ are real parameters satisfying
\begin{equation}
d=12\pi rE_{fw}c. \label{14}%
\end{equation}
Note here that $E_{tot}$ is trivially lower-bounded as
\begin{equation}
E_{tot}\geq\frac{1}{48\pi}\int_{0}^{l}\left(  \frac{\partial_{x^{+}}^{2}%
F}{\partial_{x^{+}}F}\right)  ^{2}dx^{+}. \label{16}%
\end{equation}
Substituting Eq. (\ref{15}) and Eq. (\ref{14}) into the right-hand side of Eq.
(\ref{16}) yields the following inequality:
\[
E_{tot}\geq\frac{12\pi l\left(  rE_{fw}\right)  ^{2}}{1-12\pi lrE_{fw}}.
\]
Defining the energy of $E^{\prime}$ as
\[
E_{+}=\int_{-\infty}^{0}\langle\Psi|\hat{T}_{++}(x^{+})|\Psi\rangle
dx^{+}=E_{tot}+rE_{fw},
\]
we can derive the bound in Eq. (\ref{8}).

\newpage

\newpage

\begin{thebibliography}{99}
\bibitem {ref}We would like to appreciate the reviewer for pointing out the
simple model using Rindler horizon.
\bibitem {AS}After our submission of this paper, A. Almheiri and J. Sully
upload a paper \cite{AS2}, in which they also comment on the reason why
firewalls with divergent expectation value of energy momentum tensor do not
appear for a trajectory like Eq. (\ref{10}) in the moving mirror model. They
argue that the final informative radiation from the mirror play a role of a
remnant, which entangled with early Hawking-like radiation and avoid the
emergence of firewalls. However, the "remnant" is merely a radiation of the
quantum field, and nothing special. Because the final state remains pure and
is entangled enough, their explanation does not help to resolve the AMPS paradox.
\end{thebibliography}
\end{document}